\documentclass[12pt]{article}
\usepackage{epsf}

\renewcommand{\bar}{\overline}
\newcommand{\newc}{\newcommand}
\newc{\shat}{\hat{s}}
\newc{\invpb}{\,\mbox{pb}^{-1}}
\newc{\pb}{\,\mbox{pb}}
\newc{\gev}{\,\mbox{GeV}}
\newc{\mev}{\,\mbox{MeV}}
\newc{\tev}{\,\mbox{TeV}}
\newc{\fract}{\frac}
\newc{\gsim}{\lower.7ex\hbox{$\;\stackrel{\textstyle>}{\sim}\;$}}
\newc{\lsim}{\lower.7ex\hbox{$\;\stackrel{\textstyle<}{\sim}\;$}}
\newc{\beq}{\begin{equation}}
\newc{\eeq}{\end{equation}}
\newc{\bea}{\begin{eqnarray}}
\newc{\eea}{\end{eqnarray}}
\newc{\eg}{{\it e.g.}}
\newc{\ie}{{\it i.e.}}
\newc{\etal}{{\it et al}}
\newc{\eps}{\epsilon}
\def\NPB#1#2#3{Nucl. Phys. {\bf B#1} (19#2) #3}
\def\PLB#1#2#3{Phys. Lett. {\bf B#1} (19#2) #3}

\def\PRD#1#2#3{Phys. Rev. {\bf D#1} (19#2) #3}

\catcode`@=11
\long\def\@caption#1[#2]#3{\par\addcontentsline{\csname
  ext@#1\endcsname}{#1}{\protect\numberline{\csname
  the#1\endcsname}{\ignorespaces #2}}\begingroup
    \small
    \@parboxrestore
    \@makecaption{\csname fnum@#1\endcsname}{\ignorespaces #3}\par
  \endgroup}
\catcode`@=12

\begin{document}

\begin{titlepage}
\begin{flushright}
{\rm
IASSNS-HEP-97-55\\
hep-ph/9705399\\
May 1997\\
}
\end{flushright}
\vskip 2cm
\begin{center}
{\Large\bf Modified parton distributions and the HERA high-$Q^2$ 
anomaly\footnote{Research
supported in part by DOE grant DE-FG02-90ER40542, by the W.~M.~Keck
Foundation, and the Alfred P. Sloan Foundation. Email:
{\tt babu@sns.ias.edu, kolda@sns.ias.edu,
jmr@sns.ias.edu}}}
\vskip 1cm
{\large
K.S.~Babu,
Christopher Kolda\\
and\\
John March-Russell\\}
\vskip 0.4cm
{School of Natural Sciences,\\ Institute for Advanced Study,\\
Princeton, NJ 08540\\}
\end{center}
\vskip .5cm
\begin{abstract}
A recent proposal seeks to explain the anomaly in the HERA
high-$Q^2$ neutral current $e^+p$ data by modifying the parton distribution
functions at moderate-to-high-$x$ and large $Q^2$. We investigate the
consequences of this proposal for the HERA high-$Q^2$ data itself, in 
the neutral current $e^-p$ channel and especially in the charged current
$e^\pm p$ channels. We find that there are striking signatures in the 
charged current which already serve to rule out many possibilities,
including the (symmetric) intrinsic charm hypothesis. For those not ruled
out, interesting signals are predicted in $e^-p$ scattering.
\end{abstract}
\end{titlepage}
\setcounter{footnote}{0}
\setcounter{page}{1}
\setcounter{section}{0}
\setcounter{subsection}{0}
\setcounter{subsubsection}{0}


\section{Introduction}

Recently, the H1~\cite{H1} and ZEUS~\cite{zeus} collaborations at HERA
announced an anomaly at high-$Q^2$ in the $e^+p\to eX$ neutral current
(NC) channel.  Using a combined accumulated luminosity of
$34.3\invpb$ in $e^+p\to eX$ mode at $\sqrt{s}=300\gev$, the two
experiments have
observed 24 events with $Q^2>15000\gev^2$ against a Standard Model (SM)
expectation of $13.4\pm1.0$, and 6 events with $Q^2>25000\gev^2$ against
an expectation of only $1.52\pm0.18$. Furthermore, the
high-$Q^2$ events are clustered at Bjorken-$x$ values near 0.4 to 0.5.

A number of authors have presented proposals for new physics which
might explain the NC anomaly, including: 
new contact interactions~\cite{bkmrw,altarelli,contact},
$s$-channel leptoquark production~\cite{bkmrw,altarelli,leptoquark},
and R-parity violating supersymmetry~\cite{altarelli,rparity}, and
related proposals~\cite{others}.
There exist a variety of criticisms for each of the proposed ideas, from
their highly speculative nature, to very concrete flavor problems which
they all share~\cite{bkmrw}. As such, another somewhat less
speculative idea has been proposed~\cite{tung}\ in which the anomaly in the
HERA data is simply further evidence of our inability to calculate, and often 
even reliably fit, the parton probability distribution
functions (PDFs) resulting from the non-perturbative dynamics
inside the proton. That is, Kuhlmann, \etal~\cite{tung},
have suggested a way by which the behavior of the PDFs at large-$x$ can be 
modified to explain part, or all, of the HERA NC data without disrupting
the fits of the old low-$Q^2$, low-to-moderate-$x$ data. 

It is actually a simple exercise to show that increasing the parton
densities at $Q^2\sim20,000\gev^2$ and $x\sim0.4$ to 0.5 can in fact 
fit the NC data.
It is a far more complicated question whether these changes are consistent
with all other world data.
We will not consider here the validity of the claim that such a fit can be
done which is consistent with the low-$Q^2$ data. Instead we will show that
there can be dramatic consequences for the HERA data itself in the charged
current mode at high-$Q^2$. We will find that the HERA data already
rules out a number of theoretically acceptable scenarios for modifying the
PDFs to agree with the NC data.

HERA is capable of running in two modes: $e^-p$ and $e^+p$. In the former mode
H1 and ZEUS have accumulated $1.53\invpb$ of data but have observed
no statistically significant deviations from the SM. Further, the experiments
differentiate between final state $eX$ and $\nu X$, where the neutrino is
identified through its missing $p_T$.  H1 has also announced its
findings in the $e^+p\to\nu X$ charged current (CC) channel. They find
3 events at $Q^2>20000\gev^2$ with an expectation of $0.74\pm0.39$, but
no events with $Q^2>25000\gev^2$.  ZEUS has not announced its CC data as of
this date.
Although compared to the NC data, the present CC data is much sparser,
one can still conclude from it that there can be no deviations from the SM
by more than a factor of 2 or 3 in that channel. Whether there is in fact
any deviation at all is still too uncertain to say. (Explanations of the
possible CC excess involving non-SM physics have been considered
by~\cite{altbkmrCC}.) However, given the 
sizes of the effects which we are going to find, more data in $e^+p$, and 
especially $e^-p$, mode will provide very strong constraints on the
viability of this suggestion.

\section{The proposal}

The class of proposals considered by Kuhlmann, \etal~\cite{tung},
for explaining the HERA anomaly can be thought of as
variations on the so-called ``intrinsic charm'' scenario\cite{charm}
in which it is posited that there is some non-perturbative, valence
contribution to the charm quark distributions at $Q^2\to0$.
(Current fits assume that there is no valence
charm component in the proton and generate non-zero
contributions at higher $Q^2$ only through perturbative renormalization
group flow, \ie,
gluon splitting. In this scenario one can think of the valence
structure of the proton being $uudc\bar c$ versus the usual $uud$.)

Though the underlying dynamics (and motivation) may be subtle, 
the proposal itself is straightforward: increase by hand
the parton densities inside the proton at $x\sim0.5$ and large $Q^2$ to
the point that the NC cross-section of the Standard Model
matches that observed at HERA. Such an effect can arise naturally from 
non-perturbative dynamics at low $Q^2$ if the dynamics produce a narrow
``bump'' in the PDFs at low $Q^2$ and very large $x\sim1$. This bump would
migrate down to lower $x$ as one flows (through the renormalization group) 
up to higher $Q^2$. Such a structure is difficult to rule out; the data
at low $Q^2$ and very high $x$ is limited and extractions of the 
structure functions are problematic due to non-perturbative and higher-twist
effects which can be important at large $x$. Fits available now typically
use only data with $x\lsim0.8$.
The proposal in \cite{tung}\ emphasized enhancements to the
$u$-quark density rather than that of the $c$-quark, but the difference is
irrelevant from the point of view of this paper because electroweak
physics does not distinguish among the generations.
Enhancement of any or all of the parton densities
could in principle explain the HERA data.

This proposal has strong advantages
{\sl and} disadvantages. In its favor, it requires no new physics beyond the
Standard Model and thus automatically solves the flavor problem associated
with many of the new physics interpretations~\cite{bkmrw}. 
To its detriment, it invokes
non-perturbative QCD effects which are not calculable and cannot be
predicted {\sl ab initio}, and its consistency with the moderate
$Q^2$ data taken at both HERA and the Tevatron has not been fully studied. 

\section{High-$Q^2$ tests at HERA}

Putting aside any questions of how well such a proposal can really do 
at explaining the HERA NC data while
remaining consistent with all other world data, one can ask:
what are the consequences of such a suggestion
for the NC process in $e^-$ mode, and for CC processes in either mode?

The NC result is simple: in this scenario, the NC cross-sections in both
$e^-p$ and $e^+p$ modes scale by approximately the same amount. The $Z$
couplings introduce a small helicity dependence that keeps the two modes
from being exactly the same. However, the differences between the two are
less than 10\% because the photon contribution typically dominates the 
total cross-section.
Thus this proposal predicts that once HERA has
accumulated enough data in $e^-p$ mode, they will observe a NC anomaly there
as well, and it should be roughly the same size as that observed in $e^+p$.

In discussing the CC predictions, a number of complications arise,
most stemming from the fact that the $x$ and $Q^2$ dependences of the 
NC and CC cross-sections are somewhat different. Nonetheless,
we will present fairly precise heuristic arguments about the sizes of CC 
effects, which we will then check in a full numerical calculation.

One key simplification is this: in the SM, even at high-$Q^2$,
the NC scattering is largely dominated by virtual photon exchange
in the $t$-channel. This need not have been so, since at $Q^2>>m_Z$ there
is no additional kinematic suppression of the $Z$ contributions; however,
the $Z$ coupling to quarks is generally smaller than that of the
photon. For the arguments that follow, we will ignore the $Z$ contributions
and reintroduce them only when going to the full calculation in the next
section. 

For photon exchange alone, the NC cross-section at high-$Q^2$ behaves as:
\beq
\frac{1}{x}\,\frac{d\sigma_{NC}}{dx}\propto u(x)+\bar u(x)
+\frac{1}{4}\left\{d(x)+\bar d(x)\right\}+\cdots
\eeq
where $u(x)$ is the $u$-quark parton probability distribution function (PDF)
inside a proton,
$\bar u(x)$ is the $u$-antiquark
PDF, {\sl etc.}, and the ellipses represent heavier ($s$, $c$, $b$, $t$) 
quarks. The factor of $1/4$ is the relative charge-squared of 
$u$- and $d$-type quarks.

There are four orthogonal classes of changes to the PDF's that can be
considered, each corresponding to enhancing the densities of either
$u$, $\bar u$, $d$ or $\bar d$ individually. 
That we do not have to consider changes to
the charm and strange densities is clear, since the physics in question cannot
distinguish $u$ from $c$, or $d$ from $s$. (This is not a general statement
for all experiments at all $Q^2$. For example, it does not hold even at HERA
if H1 and/or ZEUS could tag prompt charm production.)
Each of these four cases leads to a distinctive CC signature.

We will parameterize the effect of an ``intrinsic quark'' component on
the PDFs by:
\beq
q(x)=q_0(x)+q_{\rm int}(x)\equiv q_0(x)\epsilon_q(x)
\eeq
where $q(x)$ is the total parton distribution,
$q_0(x)$ is the usual fit distribution (we will use the CTEQ3 set whenever
we have to make an explicit choice~\cite{cteq}, similar results would
follow from the MRS sets~\cite{mrs}), 
$q_{\rm int}(x)$ is the intrinsic component  and
$\epsilon_q(x)\geq1$ parameterizes the effects of the enhanced
component. (The $Q^2$ dependence of $q(x)$ and $\epsilon(x)$
is implicit and of little relevance to what follows since we will only 
consider scattering within a small range of $Q^2$ values; therefore
the renormalization group running with $Q^2$ can be ignored.)
For simplicity, suppose that $\epsilon_q(x)$
is exactly 1 everywhere except in a small range of $x$ centered on $x=x_0$
for which it is much greater and roughly constant (\ie, it is roughly a
top hat distribution):
\beq
\eps_q(x)=1+\eta_q\theta(x-x_0+\delta)\theta(x_0+\delta-x)
\label{epsq}
\eeq
where $\eta_q>0$, $\theta(x)$ is the usual step function, and $\delta$ is
the half-width of the top hat.
Though the numerical arguments do not rely on making this
assumption, it does make the pedagogy simpler.

To begin, suppose that the correct fit to the HERA NC data is achieved
by changing only $u(x)$ such that $x_0$, $\delta$ and $\eps_u(x_0)$
have certain fit values. This will be the canonical scenario to which we
compare all others. For example, suppose that instead of changing the $u(x)$
PDFs, we would like to do a fit for which only $d(x)$ is changed.
Then to produce the same NC cross-section that $\eps_u(x_0)$ provided in
the $u$-case, $\eps_d(x_0)$ must shift by a larger amount, though at the same
$x=x_0$. That the shift must be larger is clear, because $d$-quark effects
on the NC cross-section are suppressed both by electric charge (the 1/4)
and by the smaller $d$-quark content in the proton, $d_0(x_0)/u_0(x_0)$.
In terms of an equation, setting the shift due
to the $\eps_u$ in the one case equal to the new shift by $\eps_d$:
\beq
\eps_u(x_0)u_0(x_0)+\frac{1}{4}d_0(x_0)=u_0(x_0)+\frac{1}{4}\eps_d(x_0)
d_0(x_0),
\eeq
giving
$\eps_d=1+4(\eps_u-1)(u_0/d_0)$, with all functions evaluated at $x=x_0$.
To get the same effect by enhancing $\bar u(x)$ alone requires $\eps_{\bar u}
=1+(\eps_u-1)(u_0/\bar u_0)$ at $x=x_0$. Similarly, changing $\bar d(x)$ alone
is identical to the case of $d(x)$, but with $d\to\bar d$ in the expression
above.

What effects do these changes have on the charged current? 
In terms of the PDFs, the CC cross-sections scale as:
\beq
\frac{1}{x}\,\frac{d\sigma^+_{CC}}{dx\,dy}\propto d(x)(1-y)^2+\bar u(x)
+\cdots,
\eeq
\beq
\frac{1}{x}\,\frac{d\sigma^-_{CC}}{dx\,dy}\propto u(x) + \bar d(x)(1-y)^2
+\cdots
\eeq
where the $+[-]$ superscript denotes scattering in $e^+p[e^-p]$ mode, and
$y$ has its usual definition in deep inelastic scattering. In the
center-of-mass frame, $y=\sin^2(\vartheta/2)$ where $\vartheta$ is the $e^\pm$
scattering angle; its presence in the expressions above follows trivially
from angular momentum conservation in the $(V-A)$ 
scattering process. Note also that
for the usual PDFs at large $x$, to a very good approximation
$\sigma^+_{CC}\propto d(x)(1-y)^2$ and $\sigma^-_{CC}\propto u(x)$.

For the moment, in order to examine the effects of enhancing the PDFs
on the CC differential cross-section, we will restrict ourselves
to the region around $x=x_0$ of half-width $\delta$. We will
denote the differential cross-section in this small region
$d\sigma^\pm_{CC}(x_0)$ as a shorthand. In the first case discussed above,
in which only $u(x_0)$ is changed in response to the NC data,
the CC signal in $e^+$ mode is unchanged, while
$d\sigma^-_{CC}(x_0)$ increases by $\eps_u(x_0)$. Thus the same relative
change in the NC data will also occur in the $e^-p$ CC data, at least in the
neighborhood of $x=x_0$.
However, because the kinematic dependence on $x$ (at high-$Q^2$)
is roughly the same in $e^+u\to e^+u$ as in $e^-u\to\nu d$, the relative
scaling of the NC to CC in the $x_0$-region will hold for all $x$.
(The preceding statement is exact in the limit $Q^2\gg m_W^2$ and when the
$Z$ contribution to NC can be ignored. Since these two conditions are not
simultaneously satisfied in general, there will be important corrections
to the results of these heuristic arguments. Nonetheless, the results derived
here do not differ greatly from the exact results, as we will see.)
Thus, to have a concrete example, if $\eps_u(x_0)$ is chosen to double the
high $Q^2$ NC cross-section, it will also (approximately)
double the high-$Q^2$ CC
cross-section in $e^-$ mode, while having no effect in $e^+$ mode.

The effect on the CC is more marked in the case of using $d(x)$ to explain the
NC anomaly. As we said above, to have the same effect on the NC data as
$\eps_u$, the $\eps_d$ must be roughly $4[u_0(x_0)/d_0(x_0)]$ times bigger
than the corresponding $\eps_u$ would have been. And since to a good
approximation
\beq
d\sigma^+_{CC}(x_0)\propto\eps_d(x_0)\simeq4\frac{u_0(x_0)}{d_0(x_0)}
\eps_u(x_0),
\eeq
it will scale by the same amount. Consider again our
example, where we demanded that $\eps_u(x_0)$ double the high-$Q^2$ NC data.
For $x_0\simeq0.5$ one finds $u_0(x_0)\simeq4d_0(x_0)$. Then to fit the
same NC data, $\eps_d(x_0)$ must be $4\cdot4=16$ times larger than
$\eps_u(x_0)$. Moving this enhancement to the CC one finds that
$d\sigma^+_{CC}(x_0)$ increases by a factor of $16\eps_u(x_0)$,
while $\sigma^-_{CC}$ is unchanged. Again the explicit kinematic
dependences on $x$ are (approximately) 
the same in the CC and NC, so the overall $\sigma^+_{CC}$ will increase 
by roughly a factor of 32 in the high-$Q^2$ region.

The same arguments go through for $\bar u$ and $\bar d$. For $\bar u$,
$d\sigma^+_{CC}(x_0)$ scales by a factor
\beq 
d\sigma^+_{CC}(x_0)\propto 1+\frac{\eps_u-1}{(y-1)^2}\,\frac{u_0}{d_0}
\simeq\frac{\eps_u}{(1-y)^2}\,\frac{u_0}{d_0},
\eeq
while $\sigma^-_{CC}$ is unchanged. For explaining the HERA anomaly we 
would be interested in $Q^2\simeq2\times10^4\gev^2$ and $x\simeq
0.5$, and thus $y=Q^2/(xs)\simeq0.5$.
Then in our recurring example wherein the NC signal is doubled,
$d\sigma^+_{CC}(x_0)$ scales by a factor of $16\eps_u$, leading to an
overall scaling of $\sigma^+_{CC}$ again by 32.

In the final case, with enhanced
$\bar d(x)$, $d\sigma^-_{CC}(x_0)$ scales by a factor 
\beq
d\sigma^-_{CC}(x_0)\propto 1+4(\eps_u-1)(1-y)^2
\eeq
while $\sigma^+_{CC}$ is unchanged. In our recurring example,
with $y\simeq0.5$, we find $d\sigma^-_{CC}(x_0)$ scales by approximately
$\eps_u$, so that the full $\sigma^-_{CC}$ at high-$Q^2$ is expected to be
double the usual SM prediction.

The various scalings of the CC mode are summarized in Table~\ref{tungtable}.
\begin{table}
\centering
\begin{tabular}{|c|c|c|c|} \hline
Parton & NC Factor & CC${}^+$ Factor & CC${}^-$ Factor \\ \hline & & & \\
$u$ & $\eps_u$ & 1 & $\eps_u$ \\ & & & \\
$d$ & $1+4(\eps_u-1)\fract{u_0}{d_0}$ & $1+4(\eps_u-1)
\fract{u_0}{d_0}$ & 1 \\ & & & \\
$\bar u$ & $1+(\eps_u-1)\fract{u_0}{\bar u_0}$ & $1+\frac{\eps_u-1}{(1-y)^2}
\,\fract{u_0}{d_0}$
& 1 \\ & & & \\
$\bar d$ & $1+4(\eps_u-1)\fract{u_0}{\bar d_0}$ & 1 &
$1+4(\eps_u-1)(1-y)^2$ \\ & & & \\ \hline
\end{tabular}
\label{tungtable}
\caption{For each parton is shown the relative sizes of contributions needed
to explain the HERA NC data, as well as the predictions for the enhancement
of the CC signal, $d\sigma^\pm_{CC}(x_0)$. All quantities are implicitly
functions of $x$ and are to be evaluated at $x=x_0$. These results are derived
in the heuristic scenario discussed in the text. Note that for $\eps_u\gg1$
the scaling factors are independent of $\eps_u$.}
\end{table}
For each parton, if we assume that it alone must have its density scaled to
explain the HERA NC data, the size of the enhancement needed (relative to
the enhancement $\eps_u$ for the $u$-quark) is given in the second column.
In the third and fourth column are then shown the relative enhancements
of $d\sigma^\pm_{CC}(x_0)$ given a fit to the NC.

Some profound results can already be extracted from these simple 
considerations. If the PDF's of $(u, d, \bar u, \bar d)$ are each 
changed respectively so as 
to double the high-$Q^2$ NC data, then the $e^+$ CC data will increase by 
factors of about
(1, 32, 32, 1) while the $e^-$ CC data will increase by (2, 1, 1, 2).
Since doubling the NC cross-section at high-$Q^2$ is roughly 
consistent with the
HERA data, we can conclude that we either expect extremely large enhancements
in the $e^+$ CC signal, or none at all. The data from H1 is not consistent
with an extremely large enhancement (such as a factor of 32), so we can
conclude that: (1) changing the PDF's of a $d$-type quark or a $\bar u$-type
quark to explain the HERA NC data is inconsistent with the HERA CC data;
(2) any other explanation
involving modifications to the $u$-type or $\bar d$-type PDFs will not
lead to a CC signal in $e^+$ mode, but will have one in $e^-$ mode;
(3) if HERA sees an enhancement in the $e^+p$ CC channel
comparable to the one in the NC channel, more than one
PDF must be modified. Note that
(2) above does not preclude a small CC enhancement in $e^+$ mode; but
any such enhancement will not be enough to also explain the NC data.

\begin{figure}
\centering
\epsfxsize=5in
\hspace*{0in}
\epsffile{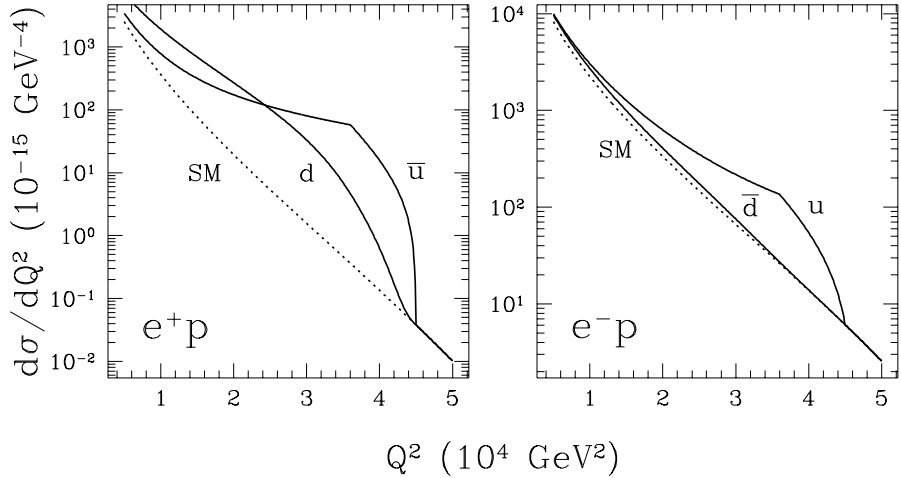}
\caption{Differential cross-section for CC $e^+p$ and $e^-p$ scattering
as a function of $Q^2$. Dotted lines are the SM prediction with unmodified
PDFs; solid lines are with modified PDFs.}
\label{figure}
\end{figure}

Up until now, our arguments have ignored the complications introduced by
keeping the $Z$ contributions in the NC, and by the non-zero $W$ mass in the
CC. To show how these results are modified in a complete calculation, we have
numerically evaluated the CC signals which would be induced by fitting to the
NC anomaly. In Figure~\ref{figure},
we have considered each of the $u$, $d$, $\bar u$ and $\bar d$ cases
separately as a solution to the HERA NC anomaly. (We are only using the H1
NC and CC data since as of this writing ZEUS has not announced their CC 
results.) We have
chosen in each case for $\eps_q(x)$ to have a top-hat form as in 
Eq.~(\ref{epsq}) with $x_0=0.45$ and $\delta=0.05$. This region in $x$
envelopes the bulk of the H1 high $Q^2$ events. The overall normalization
$\eta_q$ is chosen to provide the best possible $\chi^2$ fit to the H1 data.
(The best fit values of $\eta_q$ for $q=u,d,\bar u,\bar d$ were found to be
7.5, 95, 640, 450 using the CTEQ3M PDFs. $\eta_q$ for the sea quarks shows
a strong dependence on the PDFs used, because of the uncertainty in the sea
quark distributions at moderate $x$ and large $Q^2$. However the overall 
enhancement of the CC signal is independent of the choice.)
Then we have plotted the resulting 
$d\sigma^\pm_{CC}/dQ^2$ for each possibility.

The Figure is divided into two frames for the CC processes $e^+p\to\bar\nu X$
and $e^-p\to\nu X$ separately. For the $e^+p$ mode, the enhancement of
the CC cross-sections is marked, consistent with our heuristic derivation of
an enhancement factor of $\sim32$. For the $e^-p$ mode, the
enhancements are much smaller, once again consistent with our expectation
of factors of $\sim2$ only. 

Notice from the figure that the peak enhancements as a function
of $Q^2$ can be much larger than those of the total integrated cross-sections,
which we give in Table~\ref{fittable}. There
we have done the $Q^2$ integration in the Figure,
for $Q^2>10,000\gev^2$ and $Q^2>20,000\gev^2$.
\begin{table}
\centering
\begin{tabular}{|c|c||c|c|} \hline
PDF & Mode & $Q^2>10,000\gev^2$ & $Q^2>20,000\gev^2$ \\ \hline
$u$ & $e^-p$ & 1.9 & 2.9 \\
$d$ & $e^+p$ & 10 & 21 \\
$\bar u$ & $e^+p$ & 5.5 & 25 \\
$\bar d$ & $e^-p$ & 1.3 & 1.2 \\ \hline
\end{tabular}
\label{fittable}
\caption{Multiplicative enhancements of the CC cross-sections for $Q^2$ above
the indicated value, in the mode indicated. The PDFs in the first column
have been changed to give a best fit to the H1 NC data.}
\end{table}
The most important H1 cuts have been included in the calculation:
$y<0.9$ and $p_{T,{\rm miss}}>50\gev$. Our pedagogical derivation of the
enhancements has been shown to work reasonably well, though not exactly.
The two PDFs which affect the $e^+p$ mode both induce very large corrections
in the CC if they are invoked to explain the NC data. However the two
PDFs which affect the $e^-p$ mode induce small, but observable, corrections.
Given high statistics, HERA should be able to probe even the hardest case,
that of changing $\bar d(x)$ to explain the NC data. With the current data,
it already appears that altering $d(x)$ or $\bar u(x)$ cannot be invoked to
explain the NC anomaly.

The possibility exists of probing more complicated combinations of 
modifications because
the cross-sections at HERA are linear in the PDFs. Therefore in a scenario in
which several of the PDFs are modified so that the total modification is
a sum of individual parton modifications, weighted by some $\alpha_i$, then
the resulting CC signals are also linear combinations of those shown, again
weighted by the $\alpha_i$. For example, if enhancements to the $u$ and $d$ are
such that each one contributes half of the NC excess, then the CC signal
will be an average of the individual signals for $u$ and $d$. In particular,
the intrinsic charm scenario, for which $\eps_c=
\eps_{\bar c}$ is responsible for explaining the HERA data, the
$d\sigma^+_{CC}(x_0)$ is scaled by $1+(\eps_u-1)(u_0/2d_0)/(1-y)^2$ and
$d\sigma^-_{CC}(x_0)$ is scaled by
$(1+\eps_u)/2$ where $\eps_u=1+2(\eps_c-1)(c_0/u_0)$.
This linear behavior provides
a powerful, and simple, way to disentangle the form of the modifications
given the size of the excesses over SM in the two CC modes. Unfortunately,
with only 2 data points for 4 unknowns, one cannot deduce the full answer
using CC data alone. However, it is already enough to rule out the
intrinsic charm scenario as an explanation of the NC data, given the
numbers in Table~\ref{fittable}.

\section{Conclusions}

Modifying the PDFs to explain the HERA $e^+p$ NC data implies
an immediate modification (by about the same relative size) of
the $e^-p$ NC data,
and also implies striking patterns of modification in the CC data.
Such modifications should be easy to observe given the current size of
the NC anomaly. Further, they already rule out the intrinsic charm scenario
(with equal modifications to $c(x)$ and $\bar c(x)$),
or any other attempt to use modifications of $d$-type or $\bar u$-type
distribution functions to account for the anomaly in the HERA NC data.
Our results for the CC channel are summarized in Table~\ref{fittable}
and Figure~\ref{figure}.

If an excess in the CC signal is detected at HERA in the $e^+p$ data
of the size currently suggested by H1, then one must go to a scenario in
which more than one PDF are modified, but such that one dominates the 
contribution to the $e^+p$ NC excess. This will lead to an interesting 
signal at HERA in the $e^-p$ CC mode. In particular,
once significant luminosity has been collected in $e^-p$ mode at HERA, one 
should also be able to probe solutions to the NC anomaly which involve
only changing the distributions of $u$- and $\bar d$-type quarks.

\section*{Acknowledgments}

We wish to thank J.~Ellis, E.~Weinberg, F.~Wilczek and C.P.~Yuan
for useful conversations.

\end{document}